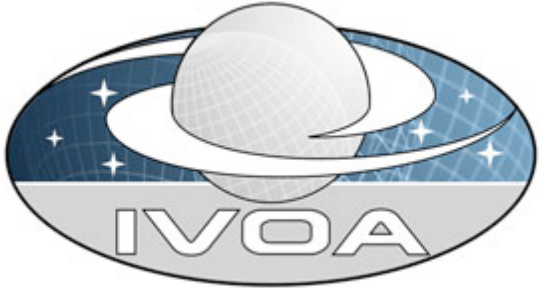

*I*nternational

*V*irtual

*O*bservatory

*A*lliance

# VOSpace specification
# Version 1.15
## *IVOA Recommendation 2009 October 07*

**Author(s):**
> Matthew Graham (Editor)
> Dave Morris
> Guy Rixon

---

## Abstract

VOSpace is the IOVA interface to distributed storage. This version extends the existing VOSpace 1.0 (SOAP-based) specification to support containers, links between individual VOSpace instances, third party APIs, and a find mechanism. Note, however, that VOSpace-1.0 compatible clients will not work with this new version of the interface.

## Status of This Document

## Acknowledgements

This document derives from discussions among the Grid and Web Services working group of the IVOA. It is particularly informed by prototypes built by Matthew Graham (Caltech, NVO) and David Morris (Cambridge, Astrogrid).

This document has been developed with support from the National Science Foundation's Information Technology Research Program under Cooperative Agreement AST0122449 with the John Hopkins University, from the UK Science and Technology Facilities Council (STFC), and from the European Commission's Sixth Framework Program via the Optical Infrared Coordination Network (OPTICON).

## Conformance related definitions

The words "MUST", "SHALL", "SHOULD", "MAY", "RECOMMENDED", and "OPTIONAL" (in upper or lower case) used in this document are to be interpreted as described in IETF standard, RFC 2119 [RFC 2119].

The **Virtual Observatory (VO)** is a general term for a collection of federated resources that can be used to conduct astronomical research, education, and outreach. The **International Virtual Observatory Alliance (IVOA)** is a global collaboration of separately funded projects to develop standards and infrastructure that enable VO applications. The International Virtual Observatory (IVO) application is an application that takes advantage of IVOA standards and infrastructure to provide some VO service.

## Contents

# 1   Introduction

VOSpace is the IVOA interface to distributed storage. It specifies how VO agents and applications can use network attached data stores to persist and exchange data in a standard way.

A VOSpace web service is an access point for a distributed storage network. Through this access point, a client can:

- add or delete data objects
- manipulate metadata for the data objects
- obtain URIs through which the content of the data objects can be accessed

VOSpace does not define how the data is stored or transferred, only the control messages to gain access. Thus, the VOSpace interface can readily be added to an existing storage system.

When we speak of "a VOSpace", we mean the arrangement of data accessible through one particular VOSpace service.

Each data object within a VOSpace service is represented as a node. A useful analogy to have in mind when reading this document is that a node is equivalent to a file.

Nodes in VOSpace have unique identifiers expressed as URIs in the `vos://` scheme, as defined below.

VOSpace 1.0 [VOSpace] defined a flat, unconnected data space. VOSpace 1.1 builds on top of this and introduces the following new functionality:

- containers - this allows the grouping of data in a hierarchical fashion
- links - this allows the federation of distinct VOSpace services
- third party APIs - this allows data objects and collections to be exposed through other interfaces
- find - this offers a more extensive search capability than is provided by list with wildcard support

## 1.1  *Typical use of a VOSpace service*

A typical use case for VOSpace is uploading a local data file to a remote VOSpace service. The user will specify the name for the data file in the VOSpace (its node identifier), and any metadata (its properties) that they want to associate with it, e.g. MIME type. They will then describe the data format (the view) they want to use in uploading the file, e.g. VOTable, and the transport protocol (the protocol) that they want to employ to upload the file, e.g. HTTP PUT. This will result in a SOAP request to the service resembling this:

```
<soapenv:Envelope xmlns:soapenv="http://schemas.xmlsoap.org/soap/envelope/"
                  xmlns:xsd="http://www.w3.org/2001/XMLSchema"
                  xmlns:xsi="http://www.w3.org/2001/XMLSchema-instance">
    <soapenv:Header>
        <soapenv:Body>
            <PushToVoSpace xmlns="http://www.net.ivoa/xml/VOSpaceContract-v1.1"
                    xmlns:vost="http://www.net.ivoa/xml/VOSpaceTypes-v1.1">
                <vost:destination  uri="vos://nvo.caltech!vospace/mytable1"
                               xsi:type="vost:UnstructuredDataNodeType">
                    <vost:properties>
                        <vost:property uri="ivo://ivoa.net/vospace/core#mimetype"
                               xsi:type="vost:PropertyType">text/xml</vost:property>
                    </vost:properties>
                </vost:destination>
                <vost:transfer>
                    <vost:view uri="ivo://ivoa.net/vospace/core#votable"/>
                    <vost:protocol uri="ivo://ivoa.net/vospace/core#http-put"/>
                </vost:transfer>
            </PushToVoSpace>
        </soapenv:Body>
    </soapenv:Header>
</soapenv:Envelope>
```

The service will reply with the representation of the data file in the VOSpace (a description of the node) and the details of the data transfer, i.e. the URL that the user will PUT the data file to. This will involve a SOAP response similar to:

```
<soapenv:Envelope xmlns:soapenv="http://schemas.xmlsoap.org/soap/envelope/"
                  xmlns:xsd="http://www.w3.org/2001/XMLSchema"
                  xmlns:xsi="http://www.w3.org/2001/XMLSchema-instance">
    <soapenv:Header>
        <soapenv:Body>
            <PushToVoSpaceResponse xmlns="http://www.net.ivoa/xml/VOSpaceContract-v1.1"
                        xmlns:vost="http://www.net.ivoa/xml/VOSpaceTypes-v1.1">
                <vost:destination  uri="vos://nvo.caltech!vospace/mytable1"
                               xsi:type="vost:UnstructuredDataNodeType">
                    <vost:properties>
                        <vost:property uri="ivo://ivoa.net/vospace/core#mimetype"
                               xsi:type="vost:PropertyType">text/xml</vost:property>
                    </vost:properties>
                </vost:destination>
                <vost:transfer>
                    <vost:view uri="ivo://ivoa.net/vospace/core#votable"/>
                    <vost:protocol uri="ivo://ivoa.net/vospace/core#http-put">
                        <vost:endpoint>http://nvo.caltech.edu:
```

```
7777/aabcf5-348874-9873ca-9a9ab4</vost:endpoint>
                    <vost:protocol>
                </vost:transfer>
            </PushToVoSpaceResponse>
        </soapenv:Body>
</soapenv:Envelope>
```

The user will then use a regular HTTP client to transfer (PUT) the local file to the specified endpoint. This illustrates an important point about VOSpace – it is only concerned with the management of data storage and transfer. A client negotiates the details of a data transfer with a VOSpace service but the actual transfer of bytes across a network is handled by other tools.

Similarly, when a user wants to retrieve a data file from a VOSpace service, they will specify the identifier for the data file in the VOSpace, the data format (view) they want to use in downloading the file, e.g. VOTable, and the transport protocol (the protocol) that they want to employ to download the file, e.g. HTTP GET. This will result in a SOAP request to the service resembling this:

```
<soapenv:Envelope xmlns:soapenv="http://schemas.xmlsoap.org/soap/envelope/"
                    xmlns:xsd="http://www.w3.org/2001/XMLSchema"
                    xmlns:xsi="http://www.w3.org/2001/XMLSchema-instance">
    <soapenv:Header>
        <soapenv:Body>
            <PullFromVoSpace xmlns="http://www.net.ivoa/xml/VOSpaceContract-v1.1"
                            xmlns:vost="http://www.net.ivoa/xml/VOSpaceTypes-v1.1">
                <vost:source>vos://nvo.caltech!vospace/mytable1</vost:source>
                <vost:transfer>
                    <vost:view uri="ivo://ivoa.net/vospace/core#votable"/>
                    <vost:protocol uri="ivo://ivoa.net/vospace/core#httpget"/>
                </vost:transfer>
            </PullFromVoSpace>
        </soapenv:Body>
</soapenv:Envelope>
```

The service will reply with a SOAP response similar to:

```
<soapenv:Envelope xmlns:soapenv="http://schemas.xmlsoap.org/soap/envelope/"
                    xmlns:xsd="http://www.w3.org/2001/XMLSchema"
                    xmlns:xsi="http://www.w3.org/2001/XMLSchema-instance">
    <soapenv:Header>
        <soapenv:Body>
            <PullFromVoSpaceResponse xmlns="http://www.net.ivoa/xml/VOSpaceContract-v1.1"
                            xmlns:vost="http://www.net.ivoa/xml/VOSpaceTypes-v1.1">
                <vost:transfer>
                    <vost:view uri="ivo://ivoa.net/vospace/core#votable"/>
                    <vost:protocol uri="ivo://ivoa.net/vospace/core#httpget">
                        <vost:endpoint>http://nvo.caltech.edu:7777/79f45-3ab0-4de2-bd6c-
ff016082fd</vost:endpoint>
                    <vost:protocol>
                </vost:transfer>
            </PullFromVoSpaceResponse>
        </soapenv:Body>
</soapenv:Envelope>
```
```



The user can then download the data file by pointing an HTTP client (e.g. web browser) at the specified endpoint.

## 1.2  *Document roadmap*

The rest of this document is structured as follows:

In Section 2, we specify the URI syntax for identifying data objects (nodes) in VOSpace.

In Section 3, we present the data model that underpins the VOSpace architecture. This consists of a number of data structures, which have XML representations that are used across the wire in SOAP message to and from a VOSpace service. These structures represent the data objects themselves (nodes), metadata that can be associated with a data object (properties), third-party interfaces to the data (capabilities), the data format used when transferring data objects across the wire (views), the transport protocol employed in a data transfer (protocols) and the data transfer itself (transfers).

In Section 4, we outline how access control policies are currently handled in VOSpace.

In Section 5, we detail the operations that VOSpace interface supports and exposes through its WSDL. These handle access to service-level metadata, the creation and manipulation of nodes within the VOSpace, access to node metadata (properties) and data transfer to and from the VOSpace.

# 2  VOSpace identifiers

The identifier for a node in VOSpace SHALL be a URI with the scheme `vos://`.

Such a URI SHALL have the following parts with the meanings and encoding rules defined in RFC2396 [2].

- scheme
- naming authority
- path
- (optional) query
- (optional) fragment identifier

The naming authority for a VOSpace node SHALL be the VOSpace service through which the node was created. The authority part of the URI SHALL be constructed from the IVO registry identifier [2] for that service by deleting the `ivo://` prefix and changing all forward-slash characters('/') in the resource key to exclamation marks ('!').

This is an example of a possible VOSpace identifier.

`vos://nvo.caltech!vospace/myresults/siap-out-1.vot`

- The URI scheme is `vos://`

Using a separate URI scheme for VOSpace identifiers enables clients to distinguish between IVO registry identifiers and VOSpace identifiers.



- `nvo.caltech!vospace`

is the authority part of the URI, corresponding to the IVO registry identifier

- `ivo://nvo.caltech/vospace`

This is the IVO registry identifier of the VOSpace service that contains the node.

- `/siap-out-1.vot` is the URI path

Slashes in the URI path imply a hierarchical arrangement of data: the data object identified `by` `vos://nvo.caltech!vospace/myresults/siap-out-1.vot` is within the container identified by `vos://nvo.caltech!vospace/myresults`.

All ancestors in the hierarchy MUST be resolvable as containers (*ContainerNodes)*, all the way up to the root node of the space (this precludes any system of implied hierarchy in the naming scheme for nodes with ancestors that are just logical entities and cannot be reified, e.g. the Amazon S3 system).

A VOSpace identifier is globally unique, and identifies one specific node in a specific VOSpace service.

A client SHOULD use the following procedure to resolve access to a VOSpace node from a VOSpace identifier:

- Extract the authority part of the VOSpace URI
- Convert the authority back to the IVO registry identifier of the VOSpace service by changing any '!' characters to '/' and adding the `ivo://` prefix
- Resolve the IVO registry identifier to an endpoint for the VOSpace service using the IVO resource registry
- Access the node via the endpoint using one of the web service methods defined in this standard

Given the example identifier

`vos://org.astrogrid.cam!vospace!container-6/siap-out-1.vot?foo=bar#baz`

processing the URI to resolve the VOSpace service would involve :

- Extract the authority part of the VOSpace URI
  - `org.astrogrid.cam!vospace!container-6`
- Convert the authority back to the IVO registry identifier of the VOSpace service by changing any '!' characters to '/'
  - `org.astrogrid.cam/vospace/container-6`
- Adding the `ivo://` prefix
  - `ivo://org.astrogrid.cam/vospace/container-6`
- Using the IVO registry to resolve the VOSpace service endpoint from the IVO identifier



# 3   VOSpace data model

## 3.1   *Nodes and node types*

We refer to the arrangement of data accessible through one particular VOSpace service as "a VOSpace".

Each data object within a VOSpace SHALL be represented as a node that is identified by a URI.

There are different types of nodes and the type of a VOSpace node determines how the VOSpace service stores and interprets the node data.

The types are arranged in a hierarchy, with more detailed types inheriting the structure of more generic types.

The following types are defined:

- *Node* is the most basic type
- *ContainerNode* describes a data item that can contain other data items
- *DataNode* describes a data item stored in the VOSpace
- *UnstructuredDataNode* describes a data item for which the VOSpace does not understand the data format

When data is stored and retrieved from an *UnstructuredDataNode*, the bit pattern read back SHALL be identical to that written.

- *StructuredDataNode* describes a data item for which the space understands the format and may make transformations that preserve the meaning of the data.

When data is stored and retrieved from a *StructuredDataNode*, the bit pattern returned MAY be different to the original. For example, storing tabular data from a VOTable file will preserve the tabular data, but any comments in the original XML file may be lost.

*LinkNode* describes a node that points to another node.

A *Node* MUST have the following elements:

- *uri* : the `vos://` identifier for the node , URI-encoded according to RFC2396 [2]
- *properties* : a set of metadata properties for the node
- *capabilities*: a third-party interface to a data object

In addition, a *DataNode* MUST have the following elements:

- *accepts*: a list of the views (data formats) that the node can accept
- *provides*: a list of the views (data formats) that the node can provide
- *busy*: a boolean flag to indicate that the data associated with the node cannot be accessed

The *busy* flag is used to indicate that an internal operation is in progress, and the node data is not available.



A *LinkNode* MUST have the following elements, in addition to the *Node* elements:

- *target*: the target URI, URI-encoded according to RFC2396 [2]

The link target can be a URI that points to any type of resource, including other VOSpace Nodes (within the same VOSpace service or in another service), or external resources outside VOSpace altogether.

The properties of a *LinkNode* do not propagate to the target of the *LinkNode*. One use case is to enable third-party annotations to be associated with a resource but without the resource itself getting cluttered with unnecessary metadata. In this case, the client creates a *LinkNode* pointing to the resource in question and then adds the annotations as properties of the *LinkNode*.

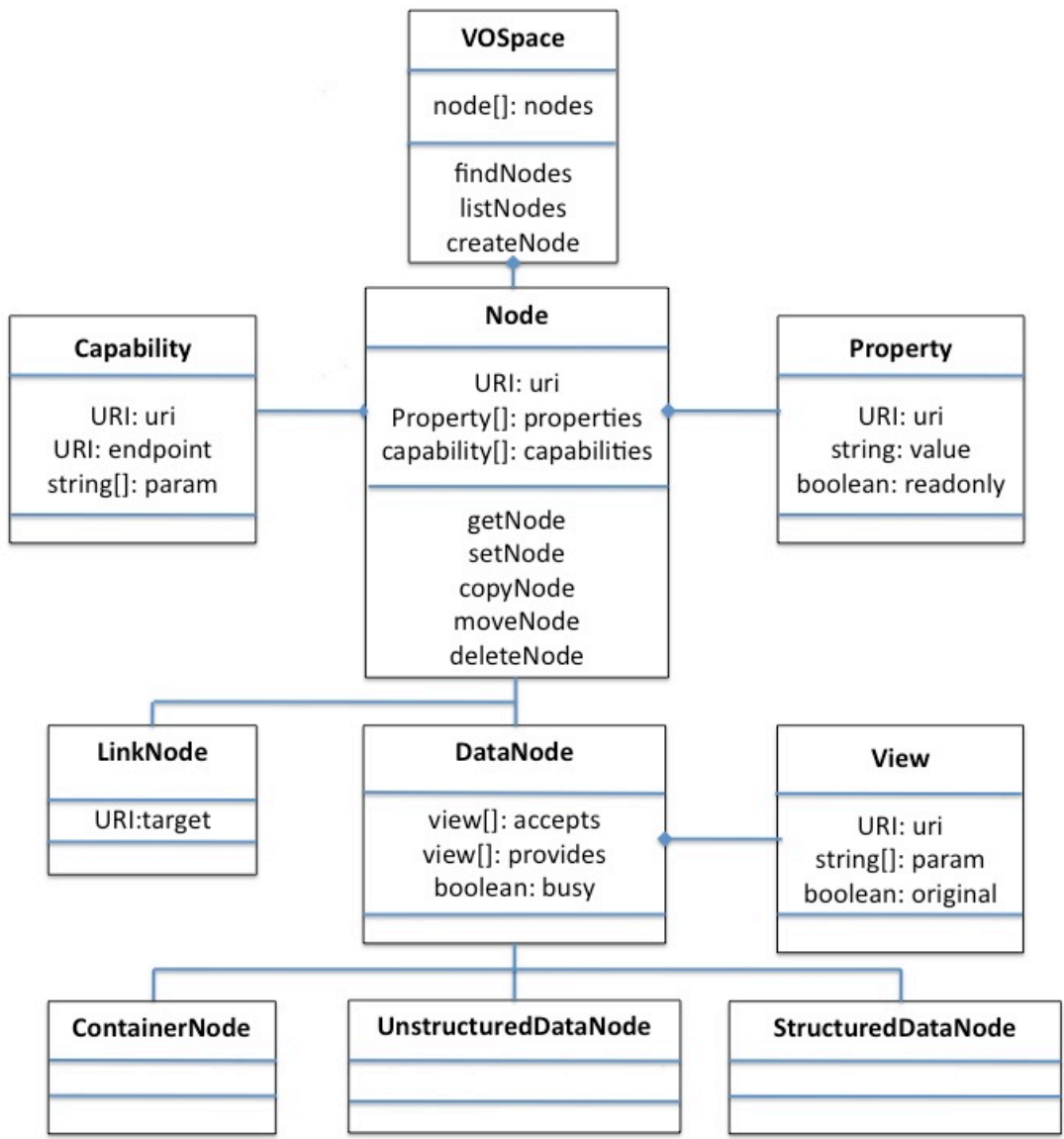



Both the *ContainerNode* and the *LinkNode* MUST have no data bytes associated with them.

The set of node types defined by this standard is closed; new types may be introduced only via new versions of the standard.

To comply with the standard, a client or service MUST be able to parse XML representations of all the node types defined in the current specification.

Note - This does not require all services to support all of the *Node* types, just that it can process an XML request containing any of the types. If the service receives a request for a type that it does not support, the service SHOULD return a *TypeNotSupported* fault. The service MUST NOT throw an XML parser error if it receives a request for a type that it does not support.

## 3.2  *Properties*

*Properties* are simple string-based metadata properties associated with a node.

Individual *Properties* should contain simple short string values, not large blocks of information. If a system needs to attach a large amount of metadata to a node, then it should either use multiple small *Properties,* or a single *Property* containing a URI or URL pointing to an external resource that contains the additional metadata.

A *Property* MUST have the following elements:

- *uri* : the *Property* identifier
- *value* : the string value of the *Property*
- *readOnly* : a boolean flag to indicate that the *Property* cannot be changed by the client

Properties may be set by the client or the service.

### 3.2.1  Property values

Unless they have special meaning to the service or client, *Properties* are treated as simple string values.

Some *Properties* may have meaning to the service, others may have meaning only to one specific type of client. A service implementation does not need to understand the meaning of all the *Properties* of a node. Any *Properties* that it does not understand can simply be stored as text strings.

### 3.2.2  Property identifiers

Every new type of *Property* SHALL require a unique URI to identify the *Property* and its meaning.

The rules for the *Property* identifiers are similar to the rules for namespace URIs in XML schema. The only restriction is that it MUST be a valid (unique) URI.

- An XML schema namespace identifier can be just a simple URN, e.g. `urn:my-namespace`
- Within the IVOA, the convention for namespace identifiers is to use a HTTP URL pointing to the namespace schema or a resource describing it

The current VOSpace schema defines *Property* identifiers as *anyURI*. The only restriction is that it



MUST be a valid (unique) URI.

- A *Property* URI can be a simple URN, e.g. `urn:my-property`

This may be sufficient for testing and development on a private system, but it is not scalable for use on a public service.

For a production system, any new *Properties* SHOULD have unique URIs that can be resolved into to a description of the *Property*.

Ideally, these should be IVO registry URIs that point to a description registered in the IVO registry:

- `ivo://my-registry/vospace/properties#my-property`

Using an IVO registry URI to identify *Properties* has two main advantages :

- IVO registry URIs are by their nature unique, which makes it easy to ensure that different teams do not accidentally use the same URI
- If the IVO registry URI points to a description registered in the IVO registry, this provides a mechanism to discover what the *Property* means

### 3.2.3  Property descriptions

If the URI for a particular *Property* is resolvable, i.e. an IVO registry identifier or a HTTP URL, then it SHOULD point to an XML resource that describes the *Property*.

A *Property* description SHOULD describe the data type and meaning of a *Property*.

A *PropertyDescription* SHOULD have the following members :

- *uri* : the formal URI of the *Property*
- *DisplayName* : A display name for the *Property*
- *Description* : A text block describing the meaning and validation rules of the *Property*

A *PropertyDescription* MAY have the following OPTIONAL members :

- *UCD* : the Universal Content Descriptor (in the UCD1+ scheme) for the *Property*
- *Unit* : the unit of measurement of the *Property*

The information in a *Property* description can be used to generate a UI for displaying and modifying the different types of *Properties.*

Note that at the time of writing, the schema for registering *PropertyDescriptions* in the IVO registry has not been finalized.

### 3.2.3.1  UI display name

If a client is unable to resolve a *Property* identifier into a description, then it may just display the identifier as a text string:

- `urn:modified-date`



If the client can resolve the *Property* identifier into a description, then the client may use the information in the description to display a human readable name and description of the *Property:*

- *Last modification date of the node data*

### 3.2.3.2  Property editors

If the client is unable to resolve a *Property* identifier into a description, or does not understand the type information defined in the description, then the client MAY treat the *Property* value as a simple text string.

If the client can resolve the *Property* identifier into a description, then the client MAY use the information in the description to display an appropriate editing tool for the *Property*.

In the current version of the specification the rules for editing *Properties* are as follows:

- A service MAY impose validation rules on the values of specific types of *Properties*
- If a client attempts to set a *Property* to an invalid value, then the service MAY reject the change
- Where possible, the validation rules for a type of *Property* SHOULD be defined in the *Property* description

Future versions of the VOSpace specification may extend the *PropertyDescription* to include more specific machine readable validation rules for a *Property* type.

Note that at the time of writing, the schema for registering validation rules in *PropertyDescriptions* has not been finalized.

## 3.2.4  Standard properties

*Property* URIs and *PropertyDescriptions* for the core set of *Properties* will be registered as soon as extension schemas are supported. However, this is not intended to be a closed list, different implementations are free to define and use their own *Properties*.

## 3.3  *Capabilities*

A *Capability* describes a third-party interface to a node. One application of this would be to enable data access to a node or its contents using a 3[rd] party service interface.

A *Capability* SHALL have the following members:

- *uri*: the *Capability* identifier
- *endpoint*: the endpoint URL to use for the third-party interface
- *param*: a set of parameters for the capability

### 3.3.1  Example use cases

A *ContainerNode* containing image files may offer a DAL SIAP capability so that the images in the container can be accessed using a SIAP service. In this way, a user could create a (DAL enabled) Container in VOSpace, transfer some images into it and then query the set of images using the SIAP interface.



Another example is a *DataNode* that provides an iRODS capability so that the data replication for this data object can be handled using the iRODS service API.

### 3.3.2 Capability identifiers

Every new type of *Capability* SHALL require a unique URI to identify the *Capability*. The rules for the *Capability* identifiers are similar to the rules for namespace URIs in XML schema. The only restriction is that it MUST be a valid (unique) URI.

- An XML schema namespace identifier can be just a simple URN, e.g. `urn:my-namespace`
- Within the IVOA, the convention for namespace identifiers is to use a HTTP URL pointing to the namespace schema, or a resource describing it.

The VOSpace schema defines *Capability* identifiers as anyURI. The only restriction is that it MUST be a valid (unique) URI.

- A Capability URI can be a simple URN, e.g. `urn:my-capability`

This may be sufficient for testing and development on a private system, but it is not suitable for use on a public service. For a production system, any new *Capabilities* SHOULD have unique URIs that can be resolved into a description of the *Capability*. Ideally, these SHOULD be IVO registry URIs that point to a description registered in the IVO registry:

- `ivo://my-registry/vospace/capabilities#my-capability`

Using an IVO registry URI to identify *Capabilities* has two main advantages:

- IVO registry URIs are by their nature unique, which makes it easy to ensure that different teams do not accidentally use the same URI
- If the IVO registry URI points to a description registered in the IVO registry, this provides a mechanism to discover how to use the Capability.

### 3.3.3 Capability descriptions

If the URI for a particular *Capability* is resolvable, i.e. an IVO registry identifier or a HTTP URL then it SHOULD point to an XML resource that describes the *Capability*.

A *CapabilityDescription* SHOULD describe the third-party interface and how it should be used in this context. A *CapabilityDescription* SHOULD have the following members:

- *uri*: the formal URI of the *Capability*
- *DisplayName*: a simple display name of the *Capability.*
- *Description*: a text block describing the third-party interface and how it should be used in this context.

Note that at the time of writing, the schema for registering *CapabilityDescriptions* in the IVO registry has not been finalized.

### 3.3.4 UI display name

If a client is unable to resolve a Capability identifier into a description then it may just display the identifier as a text string:



- Access data using `urn:edu.sdsc.irods`

If a client can resolve the *Capability* identifier into a description then the client may use the information in the description to display a human readable name and description of the *Capability*:

- Access data using iRODS

### 3.3.5 Standard capabilities

*Capability* URIs and *CapabilityDescriptions* for the core set of *Capabilities* will be registered as soon as extension schemas are supported. The following URIs SHALL be used to represent the service capabilities:

- `ivo://ivoa.net/vospace/core#vospace-1.0` SHALL be used as the capability URI for a VOSpace 1.0 service
- `ivo://ivoa.net/vospace/core#vospace-1.1` SHALL be used as the capability URI for a VOSpace 1.1 service
- `ivo://ivoa.net/vospace/core#vospace-2.0` SHALL be used as the capability URI for a VOSpace 2.0 service

If a service implementation supports more than one version of the VOSpace interface then these capability URIs can be used with a VOSpace service to identify different VOSpace capabilities for a node.

One use case for this would be a VOSpace 1.1 client talking to a service that implements both VOSpace 1.0 and VOSpace 1.1, where the client is acting on behalf of a third party agent that only understand VOSpace 1.0. In this case, the client can use the information in the VOSpace 1.0 capability to direct the third party agent to the VOSpace 1.0 endpoint.

Other standard service interfaces will also be registered, e.g.

- Cone Search
- SIAP
- SSAP
- TAP

However, this is not intended to be a closed list and different implementations are free to define and use their own *Capabilities*.

## 3.4  *Views*

A *View* describes the data formats and contents available for importing or exporting data to or from a VOSpace node.

The metadata for a Data*Node* contains two lists of *Views*.

- *accepts* is a list of *Views* that the service can accept for importing data into the *Node*
- *provides* is a list of *Views* that the service can provide for exporting data from *Node*

A *View* SHALL have the following members:



- *uri* the *View* identifier
- *original*: an optional boolean flag to indicate that the *View* preserves the original bit pattern of the data
- *param*: a set of name-value pairs that can be used to specify additional arguments for the *View*

### 3.4.1 Example use cases

#### 3.4.1.1 Simple file store

A simple VOSpace system that stores data as a binary files can just return the contents of the original file. The client supplies a *View* identifier when it imports the data, and the service uses this information to describe the data to other clients.

A file based system can use the special case identifier `'ivo://ivoa.net/vospace/core#view-any'` to indicate that it will accept any data format or *View* for a *Node*.

For example:

- A client imports a file into the service, specifying a *View* to describe the file contents
- The service stores the data as a binary file and keeps a record of the *View*
- The service can then use the *View* supplied by the client to describe the data to other clients

This type of service is not required to understand the imported data, or to verify that it contents match the *View,* it treats all data as binary files.

#### 3.4.1.2 Database store

A VOSpace system that stores data in database tables would need to be able to understand the data format of an imported file in order to parse the data and store it correctly. This means that the service can only accept a specific set of *Views,* or data formats, for importing data into the *Node*.

In order to tell the client what input data formats it can accept, the service publishes a list of specific *Views* in the *accepts* list for each *Node.*

On the output side, a database system would not be able to provide access to the original input file. The contents of file would have been transferred into the database table and then discarded. The system has to generate the output results from the contents of the database table.

In order to support this, the service needs to be able to tell the client what *Views* of the data are available.

A database system may offer access to the table contents as either VOTable or FITS files, it may also offer zip or tar.gz compressed versions of these. In which case the system needs to be able to express nested file formats such as 'zip containing VOTable' and 'tar.gz containing FITS'.

A service may also offer subsets of the data. For example, a work flow system may only want to look at the table headers to decide what steps are required to process the data. If the table contains a large quantity of data, then downloading the whole contents just to look at the header information is inefficient. To make this easier, a database system may offer a 'metadata only' *View* of the table, returning a VOTable or FITS file containing just the metadata headers and no rows.



So our example service may want to offer the following *Views* of a database table:

- Table contents as FITS
- Table contents as VOTable

- Table contents as zip containing FITS
- Table contents as zip containing VOTable

- Table contents as tar.gz containing FITS
- Table contents as tar.gz containing VOTable

- Table metadata as FITS
- Table metadata as VOTable

The service would publish this information as a list of *Views* in the *provides* section of the metadata for each *Node.*

The VOSpace specification does not mandate what *Views* a service must provide. The VOSpace specification is intended to provide a flexible mechanism enabling services to describe a variety of different *Views* of data. It is up to the service implementation to decide what *Views* of the data it can accept and provide.

### 3.4.2 View identifiers

Every new type of *View* SHALL require a unique URI to identify the *View* and its content.

The rules for the *View* identifiers are similar to the rules for namespace URIs in XML schema. The only restriction is that it MUST be a valid (unique) URI.

- An XML schema namespace identifier can be just a simple URN, e.g. `urn:my-namespace`
- Within the IVOA, the convention for namespace identifiers is to use a HTTP URL pointing to the namespace schema, or a resource describing it

The current VOSpace schema defines *View* identifiers as *anyURI*. The only restriction is that it MUST be a valid (unique) URI.

- A *View* URI can be a simple URN, e.g. `urn:my-view`

This may be sufficient for testing and development on a private system, but it is not scalable for use on a public service.

For a production system, any new *Views* SHOULD have unique URIs that can be resolved into to a description of the *View.*

Ideally, these should be IVO registry URIs that point to a description registered in the IVO registry:

- `ivo://my-registry/vospace/views#my-view`

Using an IVO registry URI to identify *Views* has two main advantages :



- IVO registry URIs are by their nature unique, which makes it easy to ensure that different teams do not accidentally use the same URI
- If the IVO registry URI points to a description registered in the IVO registry, this provides a mechanism to discover what the *View* contains

### 3.4.3 View descriptions

If the URI for a particular *View* is resolvable, i.e. an IVO registry identifier or a HTTP URL, then it SHOULD point to an XML resource that describes the *View*.

A *ViewDescription* SHOULD describe the data format and/or content of the view.

A *ViewDescription* SHOULD have the following members :

- *Uri:* the formal URI of the *View*
- *DisplayName*: A simple text display name of the *View*
- *Description*: Text block describing the data format and content of the *View*

A *ViewDescription* MAY have the following optional members :

- *MimeType* : the standard MIME type of the *View,* if applicable

However, at the time of writing, the schema for registering *ViewDescriptions* in the IVO registry has not been finalized.

#### 3.4.3.1 UI display name

If a client is unable to resolve a *View* identifier into a description, then it MAY just display the identifier as a text string:

- *Download as* `urn:table.meta.fits`

If the client can resolve the *View* identifier into a description, then the client MAY use the information in the description to display a human readable name and description of the *View*:

- *Download table metadata as FITS header*

#### 3.4.3.2 MIME types

If a VOSpace service provides HTTP access to the data contained in a *Node*, then if the *ViewDescription* contains a *MimeType* field, this SHOULD be included in the appropriate header field of the HTTP response.

### 3.4.4 Default views

The following standard URIs SHALL be used to refer to the default import and export views:

- `ivo://ivoa.net/vospace/core#anyview` SHALL be used as the view URI to indicate that a service will accept any view for an import operation
- `ivo://ivoa.net/vospace/core#binaryview` SHALL be used as the view URI to import or export data as a binary file



- `ivo://ivoa.net/vospace/core#defaultview` SHALL be used by a client to indicate that the service should choose the most appropriate view for a data export

### 3.4.4.1 Default import view

If a client imports data using this view, the data SHALL be treated as a binary BLOB, and stored as is with no additional processing. This is equivalent to the `application/binary` MIME type.

Note, this view is OPTIONAL, and the service may throw a *ViewNotSupported* exception if it does not accept this view. In particular, this view cannot be used to import data into a *StructuredDataNode* as the service needs to know about and understand the data format to be able to create the *StructuredDataNode*.

Note, this view is only valid for the data import operations, *pullToVoSpace* and *pushToVoSpace*. If this view is requested in an export operation, *pullFromVoSpace* and *pushToVoSpace*, then the service SHOULD throw a *ViewNotSupported* exception.

### 3.4.4.2 Default export view

If a client requests data using this view, the server SHALL choose whichever of the available views (the server) thinks is the most appropriate, based on how the data is stored. In a simple file-based system, this will probably be the same format that the data was originally imported in. In a database table system, this will probably either be VOTable or CVS, depending on the level of metadata available.

Note, this view is OPTIONAL, and the server may throw a *ViewNotSupported* exception if it does not provide this view. However, in most cases, it is expected that a service would be able to select an appropriate 'default' format for data held
within the service.

Note, this view is only valid for the data export operations, *pullFromVoSpace* and *pushFromVoSpace*. If this view is requested in an import operation, *pullToVoSpace* and *pushToVoSpace*, then the service SHOULD throw a *ViewNotSupported* fault.

## 3.4.5 Container views

In VOSpace 1.1, a view of a ContainerNode describes the data representation (format) of a file or data stream that represents the combined contents of the node and its children. If the view describes an archive format (tar, zip, etc.) then a service that accepts this view (format) for a ContainerNode SHALL provide access to the archive contents as children nodes of the container. Whether or not the service actually unpacks the archive is implementation dependent but the service SHALL behave as though it has done so. For example, a client may want to upload a tar file containing several images to a VOSpace service. If they associate it with (upload it to) a *UnstructuredDataNode* then it will be treated as a blob and its contents will be not be available. However, if they upload it to a *ContainerNode* with an accepts view of "tar" then the image files within the tar file will be represented as children nodes of the *ContainerNode* and accessible like any other data object within the space.

If a service provides an archive format (tar, zip, etc.) view of a ContainerNode then the service SHALL package the contents of the container and all its child nodes in the specified format.



## 3.5  *Protocols*

A *Protocol* describes the parameters required to perform a data transfer using a particular protocol.

A *Protocol* SHALL have the following members:

- *uri* : the *Protocol* identifier
- *endpoint* : the endpoint URL to use for the data transfer
- *param* : A list of name-value pairs that specify any additional arguments required for the transfer

### 3.5.1  Protocol identifiers

Every new type of *Protocol* requires a unique URI to identify the *Protocol* and how to use it.

The rules for the *Protocol* identifiers are similar to the rules for namespace URIs in XML schema. The only restriction is that it MUST be a valid (unique) URI

- An XML schema namespace identifier can be just a simple URN, e.g. `urn:my-namespace`
- Within the IVOA, the convention for namespace identifiers is to use a HTTP URL pointing to the namespace schema, or a resource describing it

The current VOSpace schema defines *Protocol* identifiers as *anyURI*. The only restriction is that it MUST be a valid (unique) URI.

- A *Protocol* URI can be a simple URN, e.g. `urn:my-protocol`

This may be sufficient for testing and development on a private system, but it is not scalable for use on a public service.

For a production system, any new *Protocols* SHOULD have unique URIs that can be resolved into to a description of the *Protocol*.

Ideally, these should be IVO registry URIs that point to a description registered in the IVO registry :

- `ivo://my-registry/vospace/protocols#my-protocol`

Using an IVO registry URI to identify *Protocols* has two main advantages:

- IVO registry URIs are by their nature unique, which makes it easy to ensure that different teams do not accidentally use the same URI
- If the IVO registry URI points to a description registered in the IVO registry, this provides a mechanism to discover how to use the *Protocol*

### 3.5.2  Protocol descriptions

If the URI for a particular *Protocol* is resolvable, i.e. an IVO registry identifier or a HTTP URL, then it SHOULD point to an XML resource that describes the *Protocol*.

A *ProtocolDescription* SHOULD describe the underlying transport protocol, and how it should be used in this context.



A *ProtocolDescription* SHOULD have the following members :

- *uri* : the formal URI of the *Protocol*
- *DisplayName* : A simple display name of the *Protocol*
- *Description* : Text block describing describing the underlying transport protocol, and how it should be used in this context

However, at the time of writing, the schema for registering *ProtocolDescriptions* in the IVO registry has not been finalized.

### 3.5.2.1 UI display name

If a client is unable to resolve a *Protocol* identifier into a description, then it MAY just display the identifier as a text string:

- *Download using* `urn:my-protocol`

If the client can resolve the *Protocol* identifier into a description, then the client MAY use the information in the description to display a human readable name and description of the *Protocol:*

- *Download using standard HTTP GET*

### 3.5.3 Standard protocols

*Protocol* URIs and *ProtocolDescriptions* for the core set of standard transport protocols will be registered as soon as extension schemas are supported. The following URIs SHALL be used to represent the standard protocols:

- `ivo://ivoa.net/vospace/core#httpget` SHALL be used as the protocol URI for a HTTP GET transfer
- `ivo://ivoa.net/vospace/core#httpput` SHALL be used as the protocol URI for a HTTP PUT transfer

However, this is not intended to be a closed list, different implementations are free to define and use their own transfer *Protocols.*

### 3.5.4 Custom protocols

There are several use cases where a specific VOSpace implementation may want to define and use a custom VOSpace transfer *Protocol,* either extending an existing *Protocol,* or defining a new one.

### 3.5.4.1 SRB gateway

One example would be a VOSpace service that was integrated with a SRB system. In order to enable the service to use the native SRB transport protocol to transfer data, the service providers would need to register a new *ProtocolDescription* to represent the SRB transport protocol.

The *ProtocolDescription* would refer to the technical specification for the SRB transport protocol, and define how it should be used to transfer data to and from the VOSpace service.



Clients that do not understand the SRB transport protocol would not recognize the URI for the SRB *Protocol,* and would ignore *Transfer* options offered using this *Protocol.*

Clients that were able to understand the SRB transport protocol would recognize the URI for the SRB *Protocol,* and could use the `'srb://..'` endpoint address in a *Protocol* option to transfer data using the SRB transport protocol.

Enabling different groups to define, register and use their own custom *Protocols* in this way means that support for new transport protocols can be added to VOSpace systems without requiring changes to the core VOSpace specification.

In this particular example, it is expected that one group within the IVOA will work with the SRB team at SDSC to define and register the *Protocol* URI and *ProtocolDescription* for using the SRB protocol to transfer data to and from VOSpace systems.
Other implementations that plan to use the SRB transport protocol in the same way could use the same *Protocol* URI and *ProtocolDescription* to describe data transfers using the SRB transport protocol.

The two implementations would then be able use the common *Protocol* URI to negotiate data transfers using the SRB transport protocol.

### 3.5.4.2  Local NFS transfers

Another example of a custom *Protocol* use case would to transfer data using the local NFS file system within an institute.

If an institute has one or more VOSpace services co-located with a number of data processing applications, all located within the same local network, then it would be inefficient to use HTTP get and put to transfer the data between the services if they could all access the same local file system.

In this case, the local system administrators could register a custom *ProtocolDescription* which described how to transfer data using their local NFS file system.

- `ivo://my.institute/vospace/protocols#internal-nfs`

Data transfers using this *Protocol* would be done using `file://` URLs pointing to locations within the local NFS file system :

- `file:///net/host/path/file`

These URLs would only have meaning to services and applications located within the local network, and would not be usable from outside the network.

Services and applications located within the local network would be configured to recognize the custom *Protocol* URI, and to use local file system operations to move files within the NFS file system.

Services and applications located outside local network would not recognize the custom *Protocol* URI and so would not attempt to use the internal file URLs to transfer data.



Note that in this example the custom *Protocol* URI and the associated *ProtocolDescription* refer to data transfers using file URLs **within a specific local NFS file system.**

If a different institute wanted to use a similar system to transfer data within their own local network, then they would have to register their own custom *Protocol* URI and associated *ProtocolDescription.*

The two different *Protocol* URIs and *ProtocolDescriptions* describe how to use the same underlying transport protocol (NFS) in different contexts.

Enabling different groups to define, register and use their own custom *Protocols* in this way means that systems can be configured to use the best available transport protocols for transferring data, without conflicting with other systems who may be using similar a transport protocol in a different context.

## 3.6  *Transfers*

A *Transfer* describes the details of a data transfer to or from a space.

A *Transfer* SHALL have the following members:

- *view* : A *View* specifying the requested *View*
    - o    For the transfer to be valid, the specified *View* must match one of those listed in the *accepts* or *provides* list of the *Node*
- *protocol* : one or more a *Protocols* specifying the transfer protocols to use
    - o    A *Transfer* may contain more than one *Protocol* element with different *Protocol* URIs
    - o    A *Transfer* may contain more than one *Protocol* element with the same *Protocol* URI with different endpoints

### 3.6.1  Synchronous transfers

Two of the VOSpace transfer methods are synchronous - the service performs the data transfer during the scope of the SOAP call.

In these methods, the client constructs a *Transfer* request containing details of the *Node* and *View* and one or more *Protocol* elements with valid endpoint addresses.

The service MAY ignore *Protocols* with URIs that it does not recognize.

If the server is unable to handle any of the requested *Protocols* in a *Transfer* request, then it SHALL respond with a fault.

The order of the *Protocols* in the request indicates the order of preference that the client would like the server to use However, this is only a suggestion, and the server is free to use its own preferences to select which *Protocol* it uses first.

The service selects the first *Protocol* it wants to use from the list and attempts to transfer the data using the *Protocol* endpoint.

If the first attempt fails, the server may choose another *Protocol* from the list and re-try the transfer using that *Protocol* endpoint.



The server may attempt to transfer the data using any or all of the *Protocols* in the list until either, the data transfer succeeds, or there are no more *Protocol* options left.

The server SHALL be allowed to only use each *Protocol* option once. This allows a data source to issue one time URLs for a *Transfer,* and cancel each URL once it has been used.

Once one of the *Protocol* options succeeds the transfer SHALL be considered to have completed, and the server is not allowed to use any of the remaining *Protocol* options. This allows a data source to issue a set of one time URLs for a transfer, and to cancel any unused URLs once the transfer has been completed.

Some *Protocols* MAY require the service to call a callback address when a data transfer completes. This behavior is specific to the *Protocol,* and SHOULD be defined in the *ProtocolDescription*.

If none of the *Protocol* options succeed, then the transfer SHALL be considered to have failed, and the service SHALL return a fault containing details of the *Protocol* options it tried.

### 3.6.2 Asynchronous transfers

Two of the VOSpace transfer methods are asynchronous - an external actor performs the data transfer outside the scope of the SOAP call.

In these methods, the client sends a template *Transfer* request to the server.

The *Transfer* request SHOULD contain details of the *Node* and *View* and one or more *Protocol* elements without endpoint addresses.

In effect, the client is sending a list of *Protocols* that it (the client) wants to use for the transfer.

The service MAY ignore *Protocols* with URIs that it does not recognize.

The service selects the *Protocols* from the request that it is capable of handling, and builds a *Transfer* response containing the selected *Protocol* elements filling in valid endpoint addresses for each of them.

The order of the *Protocol* elements in the request indicates the order of preference that the client would like to use. However, this is only a suggestion, and the server is free to use its own preferences when generating the list of *Protocols* in the response.

In effect, the server is responding with the subset of the requested *Protocols* that it (the server) is prepared to offer.

If the server is unable to accept any of the requested *Protocols,* then it SHALL respond with a fault.

On receipt of the response, the client can use the list of *Protocols* itself, or pass them on to another agent to perform the data transfer on its behalf.

The agent MAY ignore *Protocols* URIs that it does not recognize.

The agent selects the first *Protocol* it wants to use from the list and attempts to transfer the data



using the *Protocol* endpoint.

If the first attempt fails, the agent MAY choose another *Protocol* from the list and re-try the transfer using that *Protocol* endpoint.

The agent MAY attempt to transfer the data using any or all of the *Protocols* in the list until either, the data transfer succeeds, or there are no more *Protocol* options left.

The agent SHALL be allowed to only use each *Protocol* option once. This allows a data source to issue one time URLs for a *Transfer,* and cancel each URL once it has been used.

Once one of the *Protocol* options succeeds the transfer SHALL be considered to have completed, The agent is not allowed to use any of the remaining unused *Protocol* options. This allows a data source to issue a set of one time URLs for a transfer, and to cancel any unused URLs once the transfer has been completed.

Some *Protocols* MAY require the agent to call a callback address when a data transfer completes. This behavior is specific to the *Protocol,* and SHOULD be defined in the *ProtocolDescription*.

If none of the *Protocol* options succeed, then the transfer SHALL be considered to have failed.

# 4   Access control

The access control policy for a VOSpace is defined by the implementor of that space according to the use cases for which the implementation is intended.

A human-readable description of the implemented access policy must be declared in the registry metadata for the space.

These are the most probable access policies:

- No access control is asserted. Any client can create, read, write and delete nodes anonymously
- No authorization is required, but clients must authenticate an identity (for logging purposes) in each request to the space
- Clients may not create or change nodes (i.e. the contents of the space are fixed by the implementation or set by some interface other than VOSpace), but any client can read any node without authentication
- Nodes are considered to be owned by the user who created them. Only the owner can operate on a node

No operations to modify the access policy (e.g. to set permissions on an individual node) are included in this standard. Later versions may add such operations.

Where the access policy requires authentication of callers, the VOSpace service SHALL support one of the authentication methods defined in the IVOA Single Sign On profile [SSO].

# 5   Web service operations

A VOSpace 1.1 service SHALL be a SOAP service with the following operations:



## 5.1 *Service metadata*

### 5.1.1 getProtocols

Get a list of the transfer *Protocols* supported by the space service.

#### 5.1.1.1 Parameters

- none

#### 5.1.1.2 Returns

- *accepts*: A list of *Protocols* that the service SHALL accept
    - o   In this context 'accepting a protocol' means that the service SHALL act as a client for the protocol
    - o   e.g. 'accepting http-get' means the service can read data from an external HTTP web server
- *provides*: A list of *Protocols* that the service SHALL provide
    - o   In this context 'providing a protocol' means that the service SHALL act as a server for the protocol
    - o   e.g. 'providing http-get' means the service can act as a http web server

#### 5.1.1.3 Faults

- The service SHALL throw an *InternalFault* fault if the operation fails

### 5.1.2 getViews

Get a list of the *Views* and data formats supported by the space service.

#### 5.1.2.1 Parameters

- none

#### 5.1.2.2 Returns

- *accepts*: A list of *Views* that the service SHALL accept
    - o   In this context 'accepting a view' means that the service SHALL receive input data in this format
    - o   A simple file based system may use the reserved *View* URI `ivo://net.ivoa.vospace/views/any` to indicate that it can accept data in any format
- *provides*: A list of *Views* that the service SHALL provide
    - o   In this context 'providing a view' means that the service SHALL produce output data in this format
    - o   A simple file based system may use the reserved *View* URI `ivo://net.ivoa.vospace/views/any` to indicate that it can provide data in any format

#### 5.1.2.3 Faults

- The service SHALL throw an *InternalFault* fault if the operation fails



### 5.1.3 getProperties

#### 5.1.3.1 Parameters

- none

#### 5.1.3.2 Returns

- *accepts*: A list of identifiers for the *Properties* that the service SHALL accept and understand. This refers to metadata (*Properties*) that is implementation dependent but can be used by a client to control operational aspects of the service: for example, a VOSpace implementation might allow individual users to control the permissions on data objects via a *Property* called "permissions". If the VOSpace receives a data object with this *Property* then it 'understands' what this property refers to and can deal with it accordingly.
- *provides*: A list of identifiers for the *Properties* that the service SHALL provide
- *contains*: A list of identifiers for all the *Properties* currently used by *Nodes* within the service

#### 5.1.3.3 Faults

- The service SHALL throw an *InternalFault* fault if the operation fails

## 5.2  *Creating and manipulating data nodes*

### 5.2.1  createNode

Create a new node at a specified location.

#### 5.2.1.1 Parameters

- *node*: A template *Node* (as defined in Section 3.1) for the node to be created

A valid *uri* attribute SHALL be required. The name `.auto` is a reserved URI to indicate an auto-generated URI for the destination, i.e. given the following URI `vos://service/path/.auto` a service SHALL create a new unique URI for the node within `vos://service/path.`

If the *Node* `xsi:type` is not specified then a generic node of type *Node* is implied.

The permitted values of `xsi:type` are:

- `vos:Node`
- `vos:DataNode`
- `vos:UnstructuredDataNode`
- `vos:StructuredDataNode`
- `vos:ContainerNode`
- `vos:LinkNode`

When creating a new *Node* the service MAY substitute a valid subtype, i.e. If `xsi:type` is set to `vos:DataNode` then this may be implemented as a *DataNode, StructuredDataNode* or an *UnstructuredDataNode*.

The properties of the new *Node* can be set by adding *Properties* to the template. Attempting to set a *Property* that the service considers to be 'readonly' SHALL cause a



*PermissionDenied* fault.

The *accepts* and *provides* lists of *Views* for the *Node* cannot be set using this method.

The capabilities list for the *Node* cannot be set using this method.

### 5.2.1.2 Returns

- *node*: details of the new *Node*

The *accepts* list of *Views* for the *Node* SHALL be filled in by the service based on service capabilities.

The *provides* list of *Views* for the *Node* MAY not be filled in until some data has been imported into the *Node*.

The *capabilities* list for the *Node* MAY not be filled in until some data has been imported into the *Node*.

### 5.2.1.3 Faults

- The service SHALL throw an *InternalFault* fault if the operation fails
- The service SHALL throw a *DuplicateNode* fault if a *Node* already exists with the same URI
- The service SHALL throw an *InvalidURI* fault if the requested URI is invalid
- The service SHALL throw a *TypeNotSupported* fault if the type specified in `xsi:type` is not supported
- The service SHALL throw a *PermissionDenied* fault if the user does not have permissions to perform the operation
- If a parent node in the URI path does not exist then the service MUST throw a *ContainerNotFound* fault.
    - o  For example, given the URI path `/a/b/c`, the service must throw a *ContainerNotFound* fault if either `/a` or `/a/b` do not exist.
- If a parent node in the URI path is a *LinkNode*, the service MUST throw a *LinkFound* fault.
    - o  For example, given the URI path `/a/b/c`, the service must throw a *LinkFound* fault if either `/a` or `/a/b` are *LinkNode*s.

## 5.2.2 deleteNode

Delete a node.

When the target is a *ContainerNode*, all its children (the contents of the container) SHALL also be deleted.

### 5.2.2.1 Parameters

- *target*: the URI of an existing *Node*

### 5.2.2.2 Returns

- *void*

### 5.2.2.3 Faults

- The service SHALL throw an *InternalFault* fault if the operation fails
- The service SHALL throw a *PermissionDenied* fault if the user does not have permissions to



perform the operation
- The service SHALL throw a *NodeNotFound fault* if the target node does not exist
- If a parent node in the URI path does not exist then the service MUST throw a *ContainerNotFound* fault
  - o For example, given the URI path `/a/b/c`, the service must throw a *ContainerNotFound* fault if either `/a` or `/a/b` do not exist.
- If a parent node in the URI path is a *LinkNode*, the service MUST throw a *LinkFound* fault.
  - o For example, given the URI path `/a/b/c`, the service must throw a *LinkFound* fault if either `/a` or `/a/b` are *LinkNode*s.
- If the target node in the URI path does not exist, the service MUST throw a *NodeNotFound* fault.
  - o For example, given the URI path `/a/b/c`, the service must throw a *NodeNotFound* fault if `/a/b/c` does not exist.

## 5.2.3  listNodes

List nodes in a space.

In order to support large numbers of nodes, this method uses a continuation token to enable the list of results to be split across more than response.

When a target URI is a *ContainerNode*, only direct (first generation) children of the node SHALL be listed.

### 5.2.3.1  Parameters

- *token*: An optional continuation token from a previous request
  - o No token indicates a request for a new list

The server MAY impose a limited lifetime on the continuation token. If a token has expired, the server SHALL throw a fault, and the client will have to make a new request.

- *limit*: An OPTIONAL limit indicating the maximum number of results in the response
  - o No limit indicates a request for an unpaged list. However the server MAY still impose its own limit on the size of an individual response, splitting the results into more than one page if required
- *detail*: The level of detail in the returned listing
  - o `min` : The list contains the minimum detail for each *Node* with all optional parts removed – the node type should be returned
    - ▪ e.g. `<node uri="vos://service/name" xsi:type="Node"/>`
  - o `max` : The list contains the maximum detail for each *Node,* including any `xsi:type` specific extensions
  - o `properties` : The list contains a basic *node* element with a list of *properties* for each *Node* with no `xsi:type` specific extensions.
- *nodes*: A list containing zero or more *Nodes* identifying the target URIs to be listed

Only the *uri* of the *Nodes* in the request list are used, the service SHALL ignore any other elements or attributes. This method does not perform a search based on the *Node Properties* or other attributes.

The *Node* URIs in the request list MAY contain a '*' wild card character in the *name* part of



the URI (the remaining text following the last '/' character).

A single request MAY include more than one target *Node* containing a wild card. For example, the following request lists all *Nodes* with names that that match either `'*.xml'` or `'*.txt'`

```
<listNodes>
    <nodes>
        <node uri="vos://service/*.xml"/>
        <node uri="vos://service/*.txt"/>
    </nodes>
</listNodes>
```

Note that the wild card substitution for the `'*'` is a simple 'zero or more of any characters' match.

So a request for `'*.txt'` will match *Nodes* with the the following names :

- `.txt`
- `frog.txt`

Wild cards MUST only be used in the final part of the URL path: for example, a/b/c/*.txt is allowed but a/*/c/*.txt is not.

This method does not support regular expression matches.

An empty list of target *Nodes* list SHALL imply a full listing of the space.

A request with an empty list of target *Nodes* :

```
<listNodes>
    <nodes/>
</listNodes>
```

is equivalent to a request with a single target *Node* URI of `'*'` :

```
<listNodes>
    <nodes>
        <node uri="vos://service/*"/>
    </nodes>
</listNodes>
```

### 5.2.3.2 Returns

- *token* : An optional continuation token, indicating that the list is incomplete
    - o    The client MAY use this token to request the next block of *Nodes* in the sequence
    - o    No token SHALL indicate that the list is complete.



- *limit* : An OPTIONAL limit which MUST be present if a limit parameter was used in the request
  - o   If present, the value is the value from the original request and not any limit imposed by the service
- *nodes* : A list of the *Nodes* matching the requested *Node* URIs

### 5.2.3.3 Faults

- The service SHALL throw an *InternalFault* fault if the operation fails
- The service SHALL throw a *PermissionDenied* fault if the user does not have permissions to perform the operation
- The service SHALL throw a *NodeNotFound* fault if a specific target *Node* does not exist
  - o   This does not apply if the target *Node* URI contains a wild card
- The service SHALL throw an *InvalidToken* fault if it does not recognize the continuation token
- The service SHALL throw an *InvalidToken* fault if the continuation token has expired
- If a parent node in one of the search paths does not exist then the service MUST throw a *ContainerNotFound* fault.
  - o   For example, given the search path `/a/b/*`, the service must throw a *ContainerNotFound* fault if either `/a` or `/a/b` do not exist.
- If a parent node in one of the search paths is a *LinkNode*, the service MUSt throw a *LinkNode* fault.
  - o   For example, given the search path `/a/b/*`, the service must throw a *LinkFound* fault if either `/a` or `/a/b` are *LinkNode*s.

## 5.2.4 findNodes

Find nodes whose properties match the specified values.

This operation is optional.

### 5.2.4.1 Parameters

- *token*: An optional continuation token from a previous request
  - o   No token indicates a request for a new find operation.

The server MAY impose a limited lifetime on the continuation token. If a token has expired, the server SHALL throw a fault, and the client will have to make a new request.

- *limit*: An OPTIONAL limit indicating the maximum number of requests in the response
  - o   No limit SHALL indicate a request for an unpaged response. However the server MAY still impose its own limit on the size of an individual response, splitting the results into more than one page if required.
- *detail*: The level of detail in the returned response
  - o   *min*: The response contains the minimum detail for each *Node* with all optional parts removed – however, the node type SHOULD still be returned in the `xsi:type` attribute
      - e.g. `<node uri="vos://service/name" xsi:type="Node"/>`
  - o   *max*: The response contains the maximum detail for each Node, including any type specific extensions
  - o   *properties*: The response contains a basic node element with a list of properties for each *Node* with no type specific extensions
- *matches*: A list of match elements identifying the properties and values to match against and whether these should applied in conjunction (and) or disjunction (or).



The *match* element has a *uri* attribute to identify the property to which it is applying. The regular expression against which the property values are to be matched is then specified as the value of the match element:

```
<match uri="..."> regex </match>
```

The *match* elements can be combined in conjunction and/or disjunction by specifying them as subelements of `<or>` and `<and>` respectively. For example, the predicate "(property1 and property2) or property3" would be specified as:

```
<or>
    <and>
        <match uri="property1"> regex </match>
        <match uri="property2"> regex </match>
    </and>
    <match uri="property3"> regex </match>
</or>
```

The regex syntax must comply with POSIX conventions.

An empty list of `<matches>` implies a full listing of the space.

### 5.2.4.2  Returns

- *token*: An OPTIONAL continuation token, indicating that the response is incomplete
  - The client MAY use this token to request the next block of Nodes in the sequence
  - No token indicates that the list is complete
- *limit*: An OPTIONAL limit which MUST be present if a limit parameter was used in the request
  - If present, the value is the value from the original request and not any limit imposed by the service
- *nodes*: A list of the *Nodes* matching the requested properties

### 5.2.4.3  Faults

- The service SHALL throw an *InternalFault* fault if the operation fails
- The service SHALL throw a *PermissionDenied* fault if the user does not have permissions to perform the operation
- The service SHALL throw a *PropertyNotFound* fault if a particular property is specified and does not exist in the space
- The service SHALL throw an *InvalidToken* fault if it does not recognize the continuation token
- The service SHALL throw an *InvalidToken* fault if the continuation token has expired

### 5.2.5  moveNode

Move a node within a VOSpace service.

Note that this does not cover moving data between two separate VOSpace services.



Moving nodes between separate VOSpace services SHOULD be handled by the client, using the import, export and delete methods.

When the source is a *ContainerNode*, all its children (the contents of the container) SHALL also be moved to the new destination.

When the destination is an existing *ContainerNode*, the source SHALL be placed under it (i.e. within the container).

### 5.2.5.1 Parameters

- *source*: The URI of an existing *Node*
- *destination*: A template describing the new Node (as defined in section 3.1).

`.auto` is a reserved URI to indicate an autogenerated URI for the destination, i.e. `vos://service/path/.auto` SHALL cause a new unique URI for the node within `vos://service/path` to be generated.

The *Properties* from the *source Node* SHALL be combined with the *Properties* from the *destination* to create the new *Node.*

The *Node* type cannot be changed using this method.

The value of the `xsi:type` attribute on the *destination* SHALL be ignored and the new *Node* SHALL inherit the same type as the original.

The *accepts* and *provides* lists of *Views* for the new *Node* cannot be set using this method.

### 5.2.5.2 Returns

- *node*: Details of the *Node* in its new location

### 5.2.5.3 Faults

- The service SHALL throw an *InternalFault* fault if the operation fails
- The service SHALL throw a *PermissionDenied* fault if the user does not have permissions to perform the operation
- The service SHALL throw a *NodeNotFound* fault if the source *Node* does not exist
- The service SHALL throw a *DuplicateNode* fault if a *Node* already exists at the destination
- The service SHALL throw an *InvalidURI* fault if a specified URI is invalid
- The service SHALL throw an *InvalidArgument* fault if a specified value is invalid
- The service SHALL throw an *InvalidArgument* fault if request attempts to change the node type.
- If a parent node in the source URI path does not exist then the service MUST throw a *ContainerNotFound* fault.
  - o For example, given the source URI path `/a/b/c`, the service must throw a *ContainerNotFound* fault if either `/a` or `/a/b` do not exist.
- If a parent node in the source URI path is a *LinkNode*, the service MUST throw a *LinkFound* fault.
  - o For example, given the source URI path `/a/b/c`, the service must throw a *LinkFound* fault if either `/a` or `/a/b` are *LinkNode*s.
- If the source node does not exist, the service MUST throw a *NodeNotFound* fault.



- o For example, given the source URI `/a/b/c`, the service must throw a *NodeNotFound* fault if `/a/b/c` does not exist.
- If a parent node in the destination URI path does not exist then the service MUST throw a *ContainerNotFound* fault.
  - o For example, given the destination URI `/x/y/z`, the service must throw a *ContainerNotFound* fault if either `/x` or `/x/y` do not exist.
- If a parent node in the destination URI is a *LinkNode*, the service MUST throw a *LinkFound* fault.
  - o For example, given the destination URI `/x/y/z`, the service must throw a *LinkFound* fault if either `/x` or `/x/y` are *LinkNode*s.
- If the destination URI points to an existing *DataNode*, the service MUST throw a *DuplicateNode* fault.
  - o For example, given the destination URI `/x/y/z`, the service must throw a *DuplicateNode* fault if `/x/y/z` is a *DataNode*.
- If the destination URI points to an existing *LinkNode*, the service MUST throw a *LinkFound* fault.
  - o For example, given the destination URI */x/y/z*, the service must throw a *LinkFound* fault of */x/y/z* is a *LinkNode*.

## 5.2.6 copyNode

Copy a node with a VOSpace service.

Note that this does not cover copying data between two separate VOSpace services.

Copying nodes between separate VOSpace services SHOULD be handled by the client, using the import and export methods.

When the source is a *ContainerNode*, all its children (the full contents of the container) SHALL get copied, i.e. this is a deep recursive copy.

When the destination is an existing *ContainerNode*, the copy SHALL be placed under it (i.e. within the container).

### 5.2.6.1 Parameters

- *source*: The URI of an existing *Node*
- *destination*: A template *Node* (as defined in section 3.1) describing the new *Node*

`.auto` is the reserved URI to indicate an auto-generated URI for the destination, i.e. `vos://service/path/.auto` SHALL cause a new unique URI for the node within `vos://service/path` to be generated.

The *Properties* from the *source Node* SHALL be combined with the *Properties* from request to create the new *Node*.

The *Node* type cannot be changed using this method.

The value of the `xsi:type` attribute on the *destination* SHALL be ignored and the new *Node* SHALL inherit the same type as the original.

The *accepts* and *provides* lists of *Views* for the new *Node* cannot be set using this method.



#### 5.2.6.2 Returns

- *node*: Details of the new *Node*

#### 5.2.6.3 Faults

- The service SHALL throw an *InternalFault* fault if the operation fails
- The service SHALL throw a *PermissionDenied* fault if the user does not have permissions to perform the operation
- The service SHALL throw a *NodeNotFound* fault if the source *Node* does not exist
- The service SHALL throw a *DuplicateNode* fault if a *Node* already exists at the destination
- The service SHALL throw an *InvalidURI* fault if a specified URI is invalid
- The service SHALL throw an *InvalidArgument* fault if a specified value is invalid
- The service SHALL throw an *InvalidArgument* fault if request attempts to change the node type.
- If a parent node in the source URI path does not exist then the service MUST throw a *ContainerNotFound* fault.
  - For example, given the source URI path `/a/b/c`, the service must throw a *ContainerNotFound* fault if either `/a` or `/a/b` do not exist.
- If a parent node in the source URI path is a *LinkNode*, the service MUST throw a *LinkFound* fault.
  - For example, given the source URI path `/a/b/c`, the service must throw a *LinkFound* fault if either `/a` or `/a/b` are *LinkNode*s.
- If the source node does not exist, the service MUST throw a *NodeNotFound* fault.
  - For example, given the source URI `/a/b/c`, the service must throw a *NodeNotFound* fault if `/a/b/c` does not exist.
- If a parent node in the destination URI path does not exist then the service MUST throw a *ContainerNotFound* fault.
  - For example, given the destination URI `/x/y/z`, the service must throw a *ContainerNotFound* fault if either `/x` or `/x/y` do not exist.
- If a parent node in the destination URI is a *LinkNode*, the service MUST throw a *LinkFound* fault.
  - For example, given the destination URI `/x/y/z`, the service must throw a *LinkFound* fault if either `/x` or `/x/y` are *LinkNode*s.
- If the destination URI points to an existing *DataNode*, the service MUST throw a *DuplicateNode* fault.
  - For example, given the destination URI `/x/y/z`, the service must throw a *DuplicateNode* fault if `/x/y/z` is a *DataNode*.
- If the destination URI points to an existing *LinkNode*, the service MUST throw a *LinkFound* fault.
  - For example, given the destination URI */x/y/z*, the service must throw a *LinkFound* fault of */x/y/z* is a *LinkNode*.

## 5.3 *Accessing metadata*

### 5.3.1 getNode

Get the details for a specific Node.

#### 5.3.1.1 Parameters

- *target*: The URI of an existing *Node*



### 5.3.1.2 Returns

- *node*: Details of the *Node*

The full expanded record for the node SHALL be returned, including any `xsi:type` specific extensions.

### 5.3.1.3 Faults

- The service SHALL throw an *InternalFault* fault if the operation fails
- The service SHALL throw a *PermissionDenied* fault if the user does not have permissions to perform the operation
- The service SHALL throw a *NodeNotFound* fault if the target *Node* does not exist
- If a parent node in the URI path does not exist then the service MUST throw a *ContainerNotFound* fault.
  - o   For example, given the URI path `/a/b/c`, the service must throw a *ContainerNotFound* fault if either `/a` or `/a/b` do not exist.
- If a parent node in the URI path is a *LinkNode*, the service MUST throw a *LinkFound* fault.
  - o   For example, given the URI path `/a/b/c`, the service must throw a *LinkFound* fault if either `/a` or `/a/b` are *LinkNode*s.
- If the target node in the URI path does not exist, the service MUST throw a *NodeNotFound* fault.
  - o   For example, given the URI path `/a/b/c`, the service must throw a *NodeNotFound* fault if `/a/b/c` does not exist.

## 5.3.2 setNode

Set the property values for a specific node.

### 5.3.2.1 Parameters

- *node*: A *Node* containing the *Node uri* and a list of the *Properties* to set (as defined in section 3.1)

This will add or update the node properties including any `xsi:type` specific elements.

The operation is the union of existing values and new ones.

- An empty value sets the value to blank.
- To delete a *Property,* set the `xsi:nill` attribute to **true**

This method cannot be used to modify the *Node* type.

This method cannot be used to modify the *accepts* or *provides* list of *Views* for the *Node*.

### 5.3.2.2 Returns

- *node*: Details of the *Node*

The full expanded record for the node SHALL be returned, including any `xsi:type` specific extensions.



### 5.3.2.3 Faults

- The service SHALL throw an *InternalFault* fault if the operation fails
- The service SHALL throw a *PermissionDenied* fault if the request attempts to modify a *readonly Property*
- The service SHALL throw a *PermissionDenied* fault if the user does not have permissions to perform the operation
- The service SHALL throw a *NodeNotFound* fault if the target *Node* does not exist
- The service SHALL throw an *InvalidArgument* fault if a specified *property* value is invalid
- If a parent node in the URI path does not exist then the service MUST throw a *ContainerNotFound* fault.
  - For example, given the URI path `/a/b/c`, the service must throw a *ContainerNotFound* fault if either `/a` or `/a/b` do not exist.
- If a parent node in the URI path is a *LinkNode*, the service MUST throw a *LinkFound* fault.
  - For example, given the URI path `/a/b/c`, the service must throw a *LinkFound* fault if either `/a` or `/a/b` are *LinkNode*s.
- If the target node in the URI path does not exist, the service MUST throw a *NodeNotFound* fault.
  - For example, given the URI path `/a/b/c`, the service must throw a *NodeNotFound* fault if `/a/b/c` does not exist.

## 5.4 Transferring data

### 5.4.1 pushToVoSpace

Request a list of URLs to send data to a VOSpace node.

This method asks the server for a list of one or more URLs that the client can use to send data to.

The data transfer is initiated by the client, after it has received the response from the VOSpace service.

The primary use case for this method is client that wants to send some data directly to a VOSpace service.

### 5.4.1.1 Parameters

- *destination*: A description of the target *Node* (as defined in section 3.1)

A valid *uri* attribute SHALL be required.

If a *Node* already exists at the target URI, then the data SHALL be imported into the existing *Node* and the *Node properties* SHALL be updated with the *Properties* in the request.

If there is no *Node* at the destination URI, then the service SHALL create a new *Node* using the *uri,* `xsi:type` and *properties* supplied in the request (see *createNode* for details).

- *transfer*: A template (as defined in section 3.5) for the data *Transfer*

The *Transfer* template contains details of the *View* and a list of the *Protocols* that the client wants to use.



The list of *Protocols* SHOULD not contain *endpoint* addresses, the service will supply the *endpoint* addresses in the response.

The service SHALL ignore any of the requested protocols that it does not understand or is unable to support.

`.auto` is the reserved URI to indicate an auto-generated URI for the destination, i.e. `vos://service/path/.auto` SHALL cause a new unique URI for the node within `vos://service/path` to be generated.

### 5.4.1.2 Returns

- *destination*: Updated details of the destination *Node*
- *transfer*: Updated details of the data *Transfer*

The service SHALL select which of the requested *Protocols* it is willing to provide and fill in the operational details for each one – this SHOULD normally include specifying the destination URL of the transfer protocol endpoint.

The service response SHOULD not include any *Protocols* that it is unable to support.

### 5.4.1.3 Faults

- The service SHALL throw an *InternalFault* fault if the operation fails
- The service SHALL throw a *PermissionDenied* fault if the user does not have permissions to perform the operation
- The service SHALL throw a *TypeNotSupported* fault if it it is unable to create a new *Node* of the requested type
- The service SHALL throw a *ViewNotSupported* fault if a *StructuredDataNode* is requested with no *View*
- The service SHALL throw a *ViewNotSupported* fault if the service does not support the requested *View*
- The service SHALL throw a *ProtocolNotSupported* fault if it supports none of the requested *Protocols*
- The service SHALL throw an *InvalidURI* fault if the *destination* URI is invalid
- The service SHALL throw an *InvalidArgument* fault if a *View* parameter is invalid
- The service SHALL throw an *InvalidArgument* fault if a *Protocol* parameter is invalid
- The service SHALL throw a *LinkFound* fault if the target path includes a link.
- The service SHALL throw a *LinkFound* fault if the target node is a link.
- The service SHALL throw a *ContainerNotFound* fault if a parent *ContainerNode* in the target path is missing.

## 5.4.2 pullToVoSpace

Import data into a VOSpace node.

This method asks the server to fetch data from a remote location.

The data transfer is initiated by the VOSpace service and transferred direct into the target *Node*.

The data source can be another VOSpace service, or a standard HTTP or FTP server.



The primary use case for this method is transferring data from one server or service to another.

### 5.4.2.1 Parameters

- *destination*: A description of the target *Node* (as defined in section 3.1)

A valid *uri* attribute SHALL be required.

If a *Node* already exists at the target URI, then the data SHALL be imported into the existing *Node* and the *Node properties* SHALL be updated with the *Properties* in the request.

If there is no *Node* at the destination URI, then the service SHALL create a new *Node* using the *uri,* `xsi:type` and *properties* supplied in the request (see *createNode* for details).

- *transfer*: Details of the *Transfer* (as defined in section 3.5)

The *Transfer* details SHOULD include the *View* and a list of one or more *Protocols* with valid *endpoints* and parameters. The endpoint SHOULD normally refer to the source URL for the transfer protocol.

### 5.4.2.2 Returns

- *destination*: Updated details of the destination *Node*

### 5.4.2.3 Faults

- The service SHALL throw an *InternalFault* fault if the operation fails
- The service SHALL throw a *PermissionDenied* fault if the user does not have permissions to perform the operation
- The service SHALL throw a *TypeNotSupported* fault if it is unable to create a new *Node* of the requested type
- The service SHALL throw a *ViewNotSupported* fault if a *StructuredDataNode* is requested with no *View*
- The service SHALL throw a *ViewNotSupported* fault if the service does not support the requested *View*
- The service SHALL throw a *ProtocolNotSupported* fault if it supports none of the requested *Protocols*
- The service SHALL throw an *InvalidURI* fault if the destination URI is invalid
- The service SHALL throw an *InvalidArgument* fault if a *View* parameter is invalid
- The service SHALL throw an *InvalidArgument* fault if a *Protocol* parameter is invalid
- The service SHALL throw a *TransferFailed* fault if the data transfer does not complete
- The service SHALL throw an *InvalidData* fault if the data does not match the *View*
- If a parent node in the URI path does not exist then the service MUST throw a *ContainerNotFound* fault.
  - For example, given the URI path `/a/b/c`, the service must throw a *ContainerNotFound* fault if either `/a` or `/a/b` do not exist.
- If a parent node in the URI path is a *LinkNode*, the service MUST throw a *LinkFound* fault.
  - For example, given the URI path `/a/b/c`, the service must throw a *LinkFound* fault if either `/a` or `/a/b` are *LinkNode*s.
- If the target node in the URI path is a *LinkNode*, the service MUST throw a *LinkFound* fault.
  - For example, given the URI path `/a/b/c`, the service must throw a *LinkFound* fault if `/a/b/c` is a *LinkNode*.



### 5.4.2.4 Notes

In VOSpace version 1.0, the transfer is synchronous, and the SOAP call does not return until the transfer has been completed.

If the *Transfer* request contains more than one *Protocol* option, then the service MAY fail over to use one or more of the options if the first one fails. The service SHOULD try each *Protocol* option in turn until one succeeds or all have been tried.

## 5.4.3  pullFromVoSpace

Request a set of URLs that the client can read data from.

The client requests access to the data in a *Node,* and the server SHALL respond with a set of URLs that the client can read the data from.

### 5.4.3.1 Parameters

- *source* : The URI of an existing *DataNode*
- *transfer* : A template for the *Transfer* (as defined in section 3.5)

The template for the *Transfer* SHOULD contain details of the *View* and a list of the *Protocols* that the client would like to use.

The list of *Protocols* SHOULD not contain *endpoint* addresses; the service SHALL supply the *endpoint* addresses in the response.

The service SHALL ignore any of the requested protocols that it does not understand or is unable to support.

### 5.4.3.2 Returns

- *transfer*: Updated details of the data *Transfer*

The service SHALL select which of the requested *Protocols* it is willing to provide and fill in the operational details for each one – this SHOULD normally include specifying the source URL of the transfer protocol endpoint.

The service response SHOULD not include any *Protocols* that it is unable to support.

### 5.4.3.3 Faults

- The service SHALL throw an *InternalFault* fault if the operation fails
- The service SHALL throw a *PermissionDenied* fault if the user does not have permissions to perform the operation
- The service SHALL throw an *InvalidURI* fault if the *source* URI is invalid
- The service SHALL throw a *NodeNotFound* fault if the source *Node* does not exist.
- The service SHALL throw a *ProtocolNotSupported* fault if it none of the requested *Protocols* are supported
- The service SHALL throw a *ViewNotSupported* fault if it does not support the requested *View*
- The service SHALL throw an *InvalidArgument* fault if a *View* parameter is invalid
- The service SHALL throw an *InvalidArgument* fault if a *Protocols* parameter is invalid
- If a parent node in the URI path does not exist then the service MUST throw a *ContainerNotFound* fault.



- o For example, given the URI path `/a/b/c`, the service must throw a *ContainerNotFound* fault if either `/a` or `/a/b` do not exist.
- If a parent node in the URI path is a *LinkNode*, the service MUST throw a *LinkFound* fault.
  - o For example, given the URI path `/a/b/c`, the service must throw a *LinkFound* fault if either `/a` or `/a/b` are *LinkNode*s.
- If the target node in the URI path is a *LinkNode*, the service MUST throw a *LinkFound* fault.
  - o For example, given the URI path `/a/b/c`, the service must throw a *LinkFound* fault if `/a/b/c` is a *LinkNode*.
- If the target node in the URI path does not exist, the service MUST throw a *NodeNotFound* fault.
  - o For example, given the URI path `/a/b/c`, the service must throw a *NodeNotFound* fault if `/a/b/c` does not exist.

### 5.4.3.4 Notes

The endpoint URLs supplied in the response SHOULD be considered as 'one shot' URLs. A VOSpace service connected to a standard web server MAY return the public URL for the data. However, a different implementation MAY create a unique single use URL specifically for this transfer.

## 5.4.4 pushFromVoSpace

Ask the server to send data to a remote location.

The client supplies a list of URLs and asks the server to send the data to the remote location.

The transfer is initiated by the server, and the data transferred direct from the server to the remote location.

### 5.4.4.1 Parameters

- *source*: The URI of an existing *DataNode*
- *transfer*: Details of the *Transfer* (as defined in section 3.5)

The *Transfer* details SHOULD include the *View* and a list of one or more *Protocols* with valid *endpoint* and *parameters*. The endpoints will normally refer to the destination URLs for the transfer protocols.

### 5.4.4.2 Returns

- void

### 5.4.4.3 Faults

- The service SHALL throw an *InternalFault* fault if the operation fails
- The service SHALL throw a *PermissionDenied* fault if the user does not have permissions to perform the operation
- The service SHALL throw an *InvalidURI* fault if the *source* URI is invalid
- The service SHALL throw a *NodeNotFound* fault if the source *Node* does not exist
- The service SHALL throw a *ProtocolNotSupported* fault if it supports none of the requested *Protocols*
- The service SHALL throw a *ViewNotSupported* fault if it does not support the requested *View*
- The service SHALL throw an *InvalidArgument* fault if a *Protocols* parameter is invalid
- The service SHALL throw a *TransferFailed* fault if the data transfer does not complete



- If a parent node in the URI path does not exist then the service MUST throw a *ContainerNotFound* fault.
    - For example, given the URI path `/a/b/c`, the service must throw a *ContainerNotFound* fault if either `/a` or `/a/b` do not exist.
- If a parent node in the URI path is a *LinkNode*, the service MUST throw a *LinkFound* fault.
    - For example, given the URI path `/a/b/c`, the service must throw a *LinkFound* fault if either `/a` or `/a/b` are *LinkNode*s.
- If the target node in the URI path is a *LinkNode*, the service MUST throw a *LinkFound* fault.
    - For example, given the URI path `/a/b/c`, the service must throw a *LinkFound* fault if `/a/b/c` is a *LinkNode*.
- If the target node in the URI path does not exist, the service MUST throw a *NodeNotFound* fault.
    - For example, given the URI path `/a/b/c`, the service must throw a *NodeNotFound* fault if `/a/b/c` does not exist.

#### 5.4.4.4 Notes

In VOSpace version 1.0, the transfer is synchronous, and the SOAP call does not return until the transfer has been completed.

If the *Transfer* request contains more than one *Protocol* then the service MAY fail over to use one or more of the options if the first one fails. The service SHOULD try each *Protocol* option in turn until one succeeds or all have been tried.

## 5.5 *Fault arguments*

Faults thrown by a VOSpace service SHALL contain the following information:

### 5.5.1 InternalFault

This is thrown with a description of the cause of the fault.

### 5.5.2 PermissionDenied

This is thrown with no arguments.

### 5.5.3 InvalidURI

This is thrown with details of the invalid URI.

### 5.5.4 NodeNotFound

This is thrown with the URI of the missing *Node*.

### 5.5.5 DuplicateNode

This is thrown with the URI of the duplicate *Node*.

### 5.5.6 InvalidToken

This is thrown with the invalid token.

### 5.5.7 InvalidArgument

This is thrown with a description of the invalid argument, including the View or Protocol URI and the



name and value of the parameter that caused the fault.

### 5.5.8 TypeNotSupported

This is thrown with the *QName* of the unsupported type.

### 5.5.9 ViewNotSupported

This is thrown with the uri of the *View.*

### 5.5.10 InvalidData

This is thrown with any error message that the data parser produced.

The fault details must contain the URI of the missing *ContainerNode*.

### 5.5.11 LinkFoundFault

The fault details must contain the full details of the *LinkNode*.

# 6 Appendix 1: Machine readable definitions

## 6.1 *WSDL*

A proposed recommendation version of the VOSpace 1.1 WSDL can be found on the IVOA VOSpace page at: http://www.ivoa.net/twiki/bin/view/IVOA/VOSpaceHome.

## 6.2 *Message schema*

A proposed recommendation version of the XML message schema for VOSpace 1.1 can be found on the IVOA VOSpace page at: http://www.ivoa.net/twiki/bin/view/IVOA/VOSpaceHome.

# 7 Appendix 2: Compliance matrix

This table summarizes the mandatory behaviour required of a fully compliant VOSpace 1.1 service, i.e. those operations denoted as SHALL and MUST occurring. Note that for faults the general condition is specified but specific details should be checked in the relevant sections.

| Item | Description | Occurs in section(s) |
|------|-------------|---------------------|
| 1 | VOSpace node identifiers are a URI with the scheme vos:// | 2 |
| 2 | The naming authority for a VOSpace node URI is the VOSpace service through which the node was created. | 2 |
| 3 | All ancestors in a node hierarchy are resolvable as containers | 2 |
| 4 | The bit pattern for data stored in an *UnstructuredDataNode* is identical for read/write operations | 3.1 |
| 5 | A *Node* has elements: *uri, properties* and *capabilities* | 3.1 |
| 6 | A *DataNode* has elements: *uri, properties, capabilities, accepts, provides* and *busy.* | 3.1 |
| 7 | A *LinkNode* has elements: *uri, properties, capabilities* and *target* | 3.1 |
| 8 | A *ContainerNode* and a *LinkNode* have no data bytes associated with them | 3.1 |
| 9 | A VOSpace service parses XML representations of all node types | 3.1 |



| 10 | A VOSpace service does not throw an XML parser error in response to requests about unsupported node types | 3.1 |
|----|----|----|
| 11 | A *Property* has elements: *uri, value* and *readonly* | 3.2 |
| 12 | URIs must be valid (and unique) | 3.2.2, 3.3.2, 3.4.2, 3.5.2, 5.2.1.1, 5.4.1.1, 5.4.2.1 |
| 13 | A *Capability* has elements: *uri, endpoint* and *param* | 3.3 |
| 14 | Standard capabilities are represented by the URIs: | 3.3.5 |
| 15 | Each *Property* is identified by a unique URI | 3.2.2 |
| 16 | Each *Capability* is identified by a unique URI | 3.3.2 |
| 17 | Each *View* is identified by a unique URI | 3.4.2 |
| 18 | Standard views are represented by the URIs: | 3.4.4 |
| 19 | Data imported with the default import *View* is treated as a binary BLOB | 3.4.4.1 |
| 20 | Data exported with the default export *View* is returned in the most appropriate format | 3.4.4.2 |
| 21 | An archive format *View* on a *ContainerNode* provides access to the archive contents as children nodes of the container. | 3.4.5 |
| 22 | An archive format *View* specified in a data export operation on a *ContainerNode* will package the contents of the container and all its child nodes | 3.4.5 |
| 23 | Each *Protocol* is identified by a unique URI | 3.5.1 |
| 24 | Standard protocols are represented by the URIs: | 3.5.3 |
| 25 | A server responds with a fault if it is unable to handle any of the requested protocols in a data transfer. | 3.6.1, 3.6.2 |
| 26 | A server uses each requested protocol only once in a data transfer | 3.6.1, 3.6.2 |
| 27 | A data transfer is complete once a specified protocol is successful | 3.6.1, 3.6.2 |
| 28 | A data transfer has failed if none of the specified protocols has been successful | 3.6.1, 3.6.2 |
| 29 | Access policies on a VOSpace service are declared in the registered metadata for the service | 4 |
| 30 | VOSpace authentication employs IVOA SSO supported methods | 4 |
| 31 | A VOSpace service supports the operations: getProtocols, getViews, getProperties, createNode, deleteNode, findNodes, moveNode, copyNode, getNode, setNode, pushToVoSpace, pullToVoSpace, pullFromVoSpace, pushFromVoSpace | 5.1.1, 5.1.2, 5.1.3, 5.2.1, 5.2.2, 5.2.3, 5.2.4, 5.2.5, 5.2.6, 5.3.1, 5.3.2, 5.4.1, 5.4.2, 5.4.3, 5.4.4 |
| 32 | *accepts* and *provides* specify entities that the service supports | 5.1.1.2, 5.1.2.2, 5.1.3.2 |
| 33 | A URI is autogenerated if the reserved URI `.auto` is used | 5.2.1.1, 5.2.5.1, 5.2.6.1, 5.4.1.1 |
| 34 | *accepts* is filled in based on service capabilities | 5.2.1.2 |
| 35 | If a container is deleted then so are its children | 5.2.2 |
| 36 | If a container is listed, only direct children are listed | 5.2.3 |
| 37 | Only the URIs are parsed in a listing request | 5.2.3.1 |
| 38 | Wild cards can only be used in the final part of an appropriate URI | 5.2.3.1 |
| 39 | A listing request with an empty target list produces a full listing of the space | 5.2.3.1 |
| 40 | If a limit parameter is used in a request, it must be appear in the | 5.2.3.2, 5.2.4.2 |



| | response | |
|---|---|---|
| 41 | Regex expressions comply with POSIX conventions | 5.2.4.1 |
| 42 | Moving a container moves its children as well | 5.2.5 |
| 43 | Importing data into an existing container puts the new data within the container | 5.2.5, 5.2.6 |
| 44 | New properties are appended to those of an existing node | 5.2.5.1, 5.2.6.1 |
| 45 | Node types are not changed when a node is moved or copied | 5.2.5.1, 5.2.6.1 |
| 46 | Copying a container copies its children as well | 5.2.6 |
| 47 | Importing data into an existing node overwrites any preexisting data (unless it is a container) | 5.4.1.1, 5.4.2.1 |
| 48 | An *InternalFault* fault is thrown if the operation fails | 5.1.1.3, 5.1.2.3, 5.1.3.3, 5.2.1.3, 5.2.2.3, 5.2.3.3, 5.2.4.3, 5.2.5.3, 5.2.6.3, 5.3.1.3, 5.3.2.3, 5.4.1.3, 5.4.2.3, 5.4.3.3, 5.4.4.3 |
| 49 | A *DuplicateNode* fault is thrown if the specified node already exists | 5.2.1.3, 5.2.5.3, 5.2.6.3 |
| 50 | A *InvalidData* fault is thrown if the data does not match the specified *View* | 5.4.2.3 |
| 51 | A *InvalidURI* fault is thrown if the requested URI is invalid | 5.2.1.3, 5.2.5.3, 5.2.6.3, 5.4.1.3, 5.4.2.3, 5.4.3.3, 5.4.4.3 |
| 52 | A *TypeNotSupported* fault is thrown if the specified type is not supported | 5.2.1.3, 5.4.1.3, 5.4.2.3 |
| 53 | A *PermissionDenied* fault is thrown if the user does not have permissions to perform the operation | 5.2.1.3, 5.2.2.3, 5.2.3.3, 5.2.4.3, 5.2.5.3, 5.2.6.3, 5.3.1.3, 5.3.2.3, 5.4.1.3, 5.4.2.3, 5.4.3.3, 5.4.4.3 |
| 54 | A *ContainerNotFound* fault is thrown if a parent node in the specified path is not found | 5.2.1.3, 5.2.2.3, 5.2.3.3, 5.2.5.3, 5.2.6.3, 5.3.1.3, 5.3.2.3, 5.4.1.3, 5.4.2.3, 5.4.3.3, 5.4.4.3 |
| 55 | A *LinkFound* fault is thrown if a *LinkNode* is found in the specified path | 5.2.1.3, 5.2.2.3, 5.2.3.3, 5.2.5.3, 5.2.6.3, 5.3.1.3, 5.3.2.3, 5.4.1.3, 5.4.2.3, 5.4.3.3, 5.4.4.3 |
| 56 | A *NodeNotFound* fault is thrown if the specified node does not exist | 5.2.2.3, 5.2.3.3, 5.2.5.3, 5.2.6.3, 5.3.1.3, 5.3.2.3, 5.4.3.3, 5.4.4.3 |



| 57 | A *InvalidToken* fault is thrown if the specified continuation token is invalid | 5.2.3.3, 5.2.4.3 |
|----|----|----|
| 58 | A *PropertyNotFound* fault is thrown if the specified property does not exist. | 5.2.4.3 |
| 59 | A *InvalidArgument* fault is thrown if a specified value is invalid. | 5.2.5.3, 5.2.6.3, 5.3.2.3, 5.4.1.3, 5.4.2.3, 5.4.3.3, 5.4.4.3 |
| 60 | A *ViewNotSupported* fault is thrown if the specified view is not supported. | 5.4.1.3, 5.4.2.3, 5.4.3.3, 5.4.4.3 |
| 61 | A *ProtocolNotSupported* fault is thrown if the specified property is not supported. | 5.4.1.3, 5.4.2.3, 5.4.3.3, 5.4.4.3 |
| 62 | A *TransferFailed* fault is thrown if the data transfer was not successful. | 5.4.2.3, 5.4.4.3 |